\documentstyle[epsfig]{mn}

\title[DNOs and QPOs in Cataclysmic Variables: VII.]
{Dwarf nova oscillations and quasi-periodic oscillations in cataclysmic variables - VII. OY Carinae 
and oscillations in dwarf novae in quiescence}
\author[Patrick A. Woudt and Brian Warner]
       {Patrick A.~Woudt$^1$\thanks{email: Patrick.Woudt@uct.ac.za},
        Brian~Warner$^{1,2}$\\
        $^1$ Department of Astronomy, University of Cape Town, Private Bag X3,
        Rondebosch 7701, South Africa\\
        $^2$ School of Physics and Astronomy, Southampton University, Highfield, 
        Southampton SO17 1BJ, UK}
\date{2009 April 28}

\begin{document}

\maketitle

\begin{abstract}
We have observed dwarf nova oscillations (DNOs) in OY Car during outburst, down through 
decline and beyond; its behaviour is similar to what we have previously seen in VW Hyi, 
making it only the second dwarf nova to have DNOs late in outburst that continue well into quiescence. 
There are also occasional examples of DNOs in deep quiescence, well away from outburst -- they 
have properties similar to those during outburst, indicating similar physical causes and 
structures. We discuss the occurrence of DNOs in other dwarf novae and conclude that DNOs during 
quiescence are more common than often supposed and exhibit properties similar to those seen in outburst.
\end{abstract}


\section{Introduction}

In this series of papers (Woudt \& Warner 2002 (Paper I); Warner \& Woudt 2002 (Paper II); 
Warner, Woudt \& Pretorius 2003 (Paper III); Warner \& Woudt 2006 (Paper IV); Pretorius, 
Warner \& Woudt 2006 (Paper V); Warner \& Pretorius 2008 (Paper VI); Woudt et al.~2009 
(Paper VIII)) we have largely concentrated on the rapid oscillations in brightness that are 
observed in cataclysmic variables (CVs) that have accretion discs with high rates of 
mass transfer ($\dot{M}$), i.e., nova-like variables and dwarf novae in outburst. These 
have tended to be the most observed systems for dwarf nova oscillations (DNOs) and 
quasi-periodic oscillations (QPOs) because they provide the highest probability of 
exhibiting them (the rich phenomenology of DNOs and QPOs has been reviewed in Warner 
(2004)). Nevertheless, in Warner (2004) it was pointed out that there are a few reports, 
often isolated observations that have not been repeated, of optical and X-Ray DNOs and 
QPOs in dwarf novae in quiescence. An example is the 41.76 s DNO in quiescence of 
V893 Sco (Paper V). In addition, the independent variety of DNOs known as longer period DNOs 
(lpDNOs) observed at optical wavelengths (Paper III) has on one occasion been seen in quiescence 
in HT Cas (Patterson 1981) and once each in AQ Eri and OY Car (Paper III); we report more 
fully on OY Car in this paper. Examples of DNOs in early quiescence in VW Hyi, were first 
noted in Paper I, where QPOs in deep quiescence were also recorded.

   The existence of DNOs and QPOs in quiescent dwarf novae is of considerable significance 
for their relevance to possible physical interpretations. For example, a belief that such DNOs 
only appear during outburst can influence the choice of model (e.g. Piro \& Bildsten 2004).

   We first give a description of the principal properties of oscillations observed during 
outbursts. A more thorough overview can be obtained from Warner (2004) and Warner \& Woudt 
(2008). Standard DNOs exist in most but not all dwarf novae during outburst, with periods $P_{\rm DNO}$ 
in the range 6 -- 40 s, varying systematically through each outburst and passing through minimum 
period at maximum luminosity. A separate kind of DNO, of longer period 
$P_{\rm lpDNO} \sim 4 \times P_{\rm DNO}$, exists (often simultaneously with DNOs) and is less 
sensitive to luminosity than the DNOs. Two types of QPO are found: those related to DNOs, 
with $P_{\rm QPO} \sim 16 \times P_{\rm DNO}$ and frequently seen to modulate the amplitude 
of the DNOs, and another kind with much longer quasi-periods, $\sim 3000$ s. In the most 
studied dwarf nova, VW Hyi, frequency doubling and tripling of DNOs is observed late in 
outburst (Paper IV) and the fundamental period of the DNOs increases to become similar to that 
of the lpDNOs, which survive into the first few days after return to quiescence 
(Warner \& Woudt 2008; Paper VIII). 

   The most worked on model used to explain the various kinds of oscillations is the Low 
Inertia Magnetic Accretor (LIMA), developed from a suggestion of Paczynski (1978) and elaborated 
in Papers II, IV and VI, which assumes magnetically controlled accretion from the inner edge 
of a disc onto a rapidly rotating equatorial accretion belt -- in essence, an intermediate 
polar (IP) structure but accreting onto a low mass surficial layer rather than the body of the 
white dwarf primary. The DNO-related QPOs are thought to be generated by a slow prograde traveling 
wave near the inner edge of the truncated disc, partly obscuring or reprocessing radiation from 
the central regions. The optical DNOs are produced as in the IP model by high energy emission 
from an accretion zone; the resulting rotating beam can be reprocessed from the traveling 
wave as well as from the surface of the disc, producing double DNOs when simultaneously present. 
The lpDNOs are purported to arise from gas accreting along field lines attached at higher 
latitudes to the general field of the primary, and are therefore related to its global rotation 
(which will vary with latitude as angular momentum flows from gas accreting at the equator).  
Both types of DNO are observed to have abrupt small changes of period, originally thought to be 
the result of feeding different magnetic field lines (Paper II) but now realized to be at least 
partly due to alternations between the direct and reprocessed DNO beams (Paper VI).

   Of importance to the LIMA model (Paper II) is the independent evidence for hot rapidly 
rotating equatorial belts in dwarf novae, obtained from HST spectra (e.g. Cheng et al.~1997; 
Sion \& Urban 2002). Of particular relevance to this paper is the observed persistence of such 
equatorial belts in some dwarf novae for most or all of the time between outbursts (Sion et al.~2004; 
Godon et al.~2004; Urban \& Sion 2006). Pandel et al.~(2005) found a rapidly rotating 
($v \sin i \sim 1350$ km s$^{-1}$)  boundary layer 4 days after outburst maximum 
and $\sim 250$ km s$^{-1}$ 42 days after maximum; the latter is compatible 
with $\le 200$ km s$^{-1}$ deduced from widths of absorption lines in the primary (Sion 1999). 

    In Section 2 we present our new observations and their analysis. In Section 3 we list what
 else is known of quiescent oscillations. Section 4 contains a general discussion of results.

\section{New observations of OY Car}

All of our observations were made at the Sutherland site of the South African Astronomical 
Observatory (SAAO), using the University of Cape Town (UCT) CCD Photometer (O'Donoghue 1995) on the 
40- and 74-inch reflectors. Observations obtained with the SAAO 30-inch reflector (and some on the 40-inch 
reflector) made use of the UCT photoelectric photometer. A log of these observations is given in Table~\ref{dno8tab1}.

\begin{table*}
 \centering
  \caption{An overview of our data archive of OY Carinae. Observations during superoutburst and quiescence.}
  \begin{tabular}{@{}lrrccccccl@{}}
\hline
   Run        & Date &  T$^{\dag}$  & Length  & $t_{in}$ & Tel. & DNO & lpDNO & QPO & Remarks \\
              &      & (d) & (h) &     (s)     &     &       &     &   \\
\hline 
 \multicolumn{10}{l}{Observations during superoutburst}\\[5pt]
 S6722 & 2003 Feb 01 &  --5.9  & 2.80  & 4,5  & 40-in   & $\surd$ &  --   & $\surd$ & DNOs at 17 -- 18 s (3 -- 7 mmag)\\
       &             &         &       &      &         &     &       &     & QPOs at 278 s (11 mmag)\\
 S6724 & 2003 Feb 01 &  --5.7  & 1.94  & 5    & 40-in   & $\surd$ &  $\surd$  & $\surd$ & DNOs at 17 -- 18 s (3 -- 6 mmag)\\
       &             &         &       &      &         &     &       &     & lpDNO at 116 s (9 mmag)\\
       &             &         &       &      &         &     &       &     & QPOs at 315 s (9 mmag) \\
 S6727 & 2003 Feb 02 &  --4.9  & 1.39  & 5,6  & 40-in   &  -- &  --   & --  & No DNOs $>$ 2 mmag \\
 S6730 & 2003 Feb 02 &  --4.8  & 1.21  & 5    & 40-in   &  -- &  --   & --  & No DNOs $>$ 2 mmag \\
 S5482$^{\ddag}$ & 1992 Apr 21 &  --2.9  & 2.25  & 5    & 40-in   &  -- &  --   & --  & No DNOs $>$ 2 mmag \\
 S6739 & 2003 Feb 04 &  --2.8  & 1.97  & 5    & 40-in   &  -- &  --   & --  & No DNOs $>$ 2 mmag \\
 S5484$^{\ddag}$ & 1992 Apr 22 &  --2.0  & 1.20  & 5    & 40-in   &  -- &  --   & --  & No DNOs $>$ 4 mmag  \\
 S6743 & 2003 Feb 05 &  --1.9  & 2.47  & 5    & 40-in   &  -- &  --   & --  & No DNOs $>$ 3 mmag \\
 S6748 & 2003 Feb 07 &    0.2  & 1.54  & 5    & 40-in   & $\surd$ &  --   & --  & DNOs at 29 s (12 mmag) \\
 S6752 & 2003 Feb 09 &    2.0  & 1.06  & 6    & 40-in   &  -- &  --   & --  & No DNOs $>$ 10 mmag\\
 S6754 & 2003 Feb 09 &    2.2  & 2.81  & 5    & 40-in   & $\surd$ &  --   & $\surd$ & DNOs at 50 -- 60 s (11 -- 12 mmag) \\
       &             &         &       &      &         &     &       &     & QPOs at 305, 571 s (22 -- 34 mmag) \\
 S6755 & 2003 Feb 10 &    3.2  & 0.58  & 6    & 40-in   & $\surd$ &  --   & --  & DNOs at 50 s (12 mmag) \\[5pt]
 \multicolumn{10}{l}{Observations during quiescence}\\[5pt]
 S2966$^{\ddag}$ & 1982 Mar 24 & $\sim$ 46$^{\star}$ &  2.02  & 10       & 30-in & \multicolumn{2}{c}{$\surd$}    & -- & (lp)DNOs at 104 s (25 mmag) \\
 S2970$^{\ddag}$ & 1982 Mar 25 & $\sim$ 47$^{\star}$ &  1.49  & 10       & 30-in & \multicolumn{2}{c}{--}     & -- & No DNOs $>$ 10 mmag \\
 S2974$^{\ddag}$ & 1982 Mar 28 & $\sim$ 50$^{\star}$ &  0.95  & 10       & 30-in & \multicolumn{2}{c}{--}     & -- & No DNOs $>$ 10 mmag \\
 S2982$^{\ddag}$ & 1982 May 18 &$\sim$ 101$^{\star}$ &  1.04  & 10       & 30-in & \multicolumn{2}{c}{$\surd$}    & -- & (lp)DNOs at 130 s (17 mmag) \\
 S3273$^{\ddag}$ & 1984 Mar 03 & $\sim$ 40$^{\star}$ &  0.25  &  1       & 30-in & \multicolumn{2}{c}{--}     & -- & No DNOs $>$ 20 mmag \\
 S3826$^{\ddag}$ & 1986 May 02 & $\sim$ 20$^{\star}$ &  0.69  & 5        & 30-in & \multicolumn{2}{c}{--}     & -- & No DNOs $>$ 20 mmag \\
 S6050 & 2000 Feb 01 &$\sim$ 202$^{\star}$ &  1.36  & 10       & 40-in & \multicolumn{2}{c}{--}     & -- & No DNOs $>$ 9 mmag  \\
 S6055 & 2000 Feb 03 &$\sim$ 204$^{\star}$ &  2.32  & 5,8      & 40-in & \multicolumn{2}{c}{--}     & -- & No DNOs $>$ 5 mmag \\
 S6485 & 2002 Feb 15 & $\sim$ 34$^{\star}$ &  6.91  & 5,7,10   & 74-in & \multicolumn{2}{c}{$\surd$}    & $\surd$ & (lp)DNOs at 48 -- 74 s (9 -- 26 mmag) \\
       &             &                    &        &          &       &             &              &     & (QPOs at 585 s (33 mmag)              \\
 S6488 & 2002 Feb 16 & $\sim$ 35$^{\star}$ &  2.88  & 7        & 74-in & \multicolumn{2}{c}{$\surd$}    & --  & (lp)DNOs at 48, 61 s (26 -- 29 mmag) \\
\hline
\end{tabular}
\newline
{\footnotesize $^{\dag}$ $T = 0$ is defined by the average superoutburst profile (see Fig.~\ref{dno8fig1}); $^{\ddag}$ data obtained with the UCT photometer; 
$^{\star}$ Time since last recorded outburst.}
\label{dno8tab1}
\end{table*}

\subsection{Observations during outburst and early quiescence}

We have previously described selected observations of DNOs in the SU UMa type dwarf nova 
OY Car (Paper III; Warner \& Woudt 2005); here we report on further light curves and analyse 
the whole data set. It emerges that the DNOs in OY Car behave in a manner similar to those 
of VW Hyi in outburst (Papers I and IV) and into quiescence, so we discuss their full evolution.

   OY Car is an eclipsing system with an orbital period of 1.51 h, discovered by Vogt (1979), 
and has normal outbursts with an (uncertain) average interval of 160 d and superoutbursts 
that recur on a timescale of 346 d (Warner 1995). These are greater intervals than for VW Hyi, 
which has 28 d and 185 d respectively, suggesting that a lower mass transfer rate is 
operating in OY Car.

\begin{figure}
\centerline{\hbox{\psfig{figure=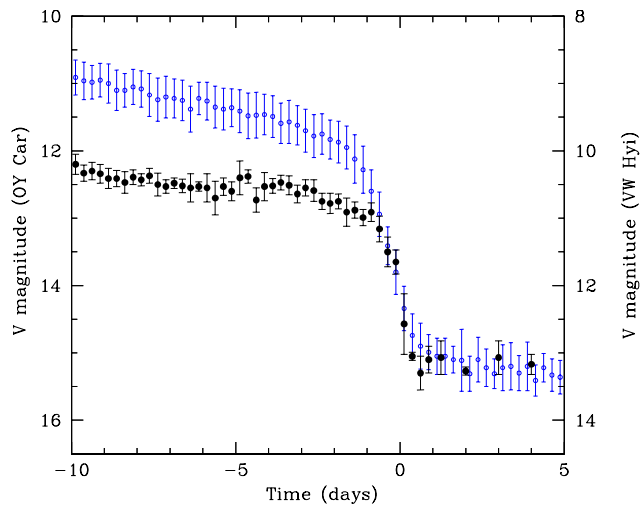,width=8.4cm}}}
  \caption{The average superoutburst light curve of OY Car (filled circles), compared to the
average superoutburst profile of VW Hyi (open circles).}
 \label{dno8fig1}
\end{figure}

   All of the outburst DNOs observed by us and by others happen to have been made during or 
shortly after superoutbursts. To provide a fiducial time during outburst with which to refer 
the DNOs (cf. the VW Hyi outburst templates in fig 5 of Paper I) we used magnitudes available 
from the American Association of Variable Star Observers to produce an average 
superoutburst light curve, shown together with one for VW Hyi in Fig.~\ref{dno8fig1} (in OY Car, $T$ = 0 
corresponds to the brightness dropping through $V$ = 14.1). Comparison of this with the American
Association of Variable Star Observers (AAVSO) 
observations for the appropriate individual outbursts enables us to set our observations and 
those of previous observers on a common evolutionary time scale. For observations made during 
quiescence we list time since the last reported outburst.

   We have computed Fourier transforms (FTs) of the light curves, subdividing them in order 
to be more sensitive to the occurrence of DNOs, which can vary in period and amplitude. When 
present we follow the oscillations using amplitude/phase diagrams ($A - \phi$), which are simply 
observed-calculated values of least squares fits of short lengths of light curve to sinusoids of 
appropriate periods. Table~\ref{dno8tab1} notes whether significant oscillations occurred during the runs and 
Table~\ref{dno8tab2} gives details for those sections, with mean periods and amplitudes determined from 
the least squares fits.

\begin{figure}
\centerline{\hbox{\psfig{figure=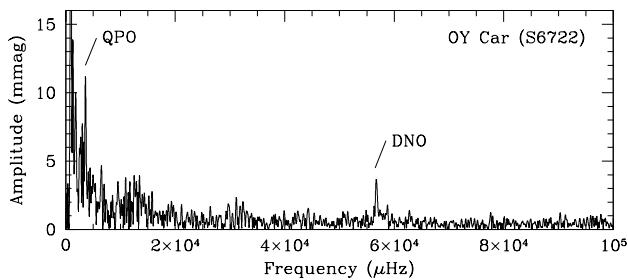,width=8.4cm}}}
  \caption{The Fourier transform of a section of run S6722. Both the QPO and DNO frequencies are
marked.}
 \label{dno8fig2}
\end{figure}

  We found oscillations in only one of the runs made near a maximum of outburst; the FT for 
the combined subsections IV, V and VI of run S6722 is shown in Fig.~\ref{dno8fig2} and shows DNOs and QPOs. 
The FT for a part of Section I of run S6724 (which was a continuation of S6722 on the same 
night) showed DNOs and lpDNos and was given in fig. 16 of Paper III. The DNOs (Table.~\ref{dno8tab2}) show 
small variations in period around $\sim 17.6$ s, typical of that seen in other dwarf novae at maximum 
of outburst. lpDNOs were evident only in S6724, and include perhaps a switch to a first harmonic; 
QPOs were present in the latter half of S6722 and the first parts of S6724. The ratios 
$P_{\rm QPO}/P_{\rm DNO} \sim 300/17.6 = 17.0$ and $P_{\rm lpDNO}/P_{\rm DNO} \sim 65/17.6 = 3.7$ are 
close to the values 16 and 4 that are typically found (see above).

\begin{figure}
\centerline{\hbox{\psfig{figure=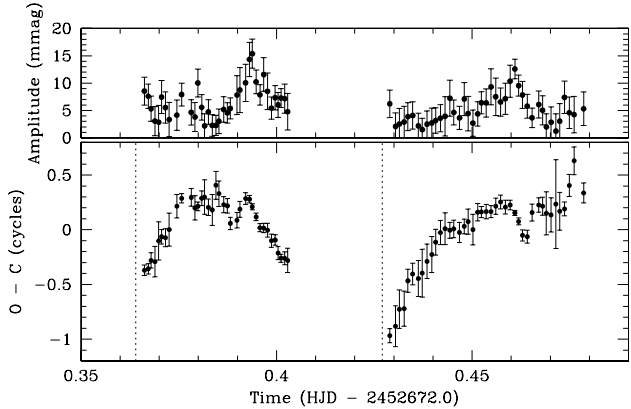,width=8.4cm}}}
  \caption{The $A - \phi$ diagram for run S6722, relative to a fixed period of 17.59 s; phase 
variations are shown in the lower panel, amplitude variations are displayed in the top panel.
The dashed vertical lines in the lower panel indicate times of mid-eclipse according to the
ephemeris of Greenhill et al.~(2006).}
 \label{dno8fig3}
\end{figure}

   The DNO $A - \phi$ diagram for run S6722 is shown in Fig.~\ref{dno8fig3} (a period of 17.59 s 
was used for comparison; during the central gap no data were taken). There is a progressive reduction 
in period after each eclipse, as seen in the individual periods listed in Table~\ref{dno8tab2} and 
the sequence of nearly straight line sections on the phase variations 
(which are themselves used to define the boundaries of the sections). The same behaviour is seen in 
the second part of this night (run S6724 -- see periods in Table~\ref{dno8tab2}). This systematic 
effect with orbital phase can be caused by the varying obscuration of an accretion disc with 
changing thickness around its rim.

   We may be considered fortunate to have detected relatively strong oscillations in one of 
only two observations made close to maximum: in VW Hyi the DNOs are very rare during maxima of 
outbursts (Paper I). We have observations on two nights during the April 1992 superoutburst of 
OY Car, during which no oscillations were detected, but on the next night Marsh \& Horne (1988) 
found 18.0 s DNOs, discuss further below.

  On only one occasion during the decline phase of an outburst (which is short lived and 
difficult to catch) have we detected DNOs: 28.9 s in run S6748 (Table~\ref{dno8tab2}). But for up to three 
days after the end of outburst oscillations of considerable amplitude appear to be quite common.

\begin{figure}
\centerline{\hbox{\psfig{figure=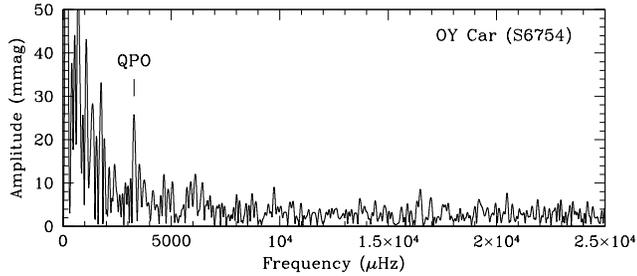,width=8.4cm}}}
  \caption{The Fourier transform of run S6754. The 305-s QPO is marked.}
 \label{dno8fig4}
\end{figure}

      Run S6754, made just after starting quiescence, has DNOs more or less continually present 
after the first $\sim 1000$ s. There are prominent double DNOs in sections I and II, which have beat 
periods near 350 s, and there is a prominent QPO visible in the FT for the whole run, at a 
period of 305 s (Fig.~\ref{dno8fig4}). Run S6755, made on the following night, has a DNO near the lower 
value of the previous DNOs.

     The lpDNOs observed simultaneously with DNOs during outburst in run S6724 serve to indicate 
the lpDNO periods. In VW Hyi we found that the lpDNOs only increased slightly in period through 
outburst and into quiescence, so by analogy we have classified the $\sim 50 - 70$ s oscillations seen 
late after outburst as lpDNOs (Table~\ref{dno8tab2}), but there is some ambiguity because we have not detected 
any ordinary DNOs in the same runs.

     There have been only two previous studies of DNOs in OY Car. Schoembs (1986) caught OY Car in 
late superoutburst and followed it into quiescence with 1 s integration times. He was the 
first to detect DNOs in this object and studied in detail oscillations with periods of 19.44, 
20.48 and 27.99 s during outburst, also finding periodicities at 55 and 120 s in quiescence a 
week later. (As can be seen in fig. 8 of Schoembs (1986) his O--C curves do not return to their 
initial phases, indicating use of comparison sine waves with periods too short; the optimum 
periods would be up to two tenths of a second longer than those quoted by Schoembs.) Marsh \& Horne 
(1988) observed with HST for $\sim 2250$ s towards the end of a superoutburst, finding a double 
DNO with periods 17.94 and 18.16 s, or an implied beat period of 1480 s; these we have discussed 
in Paper II. 

\begin{figure}
\centerline{\hbox{\psfig{figure=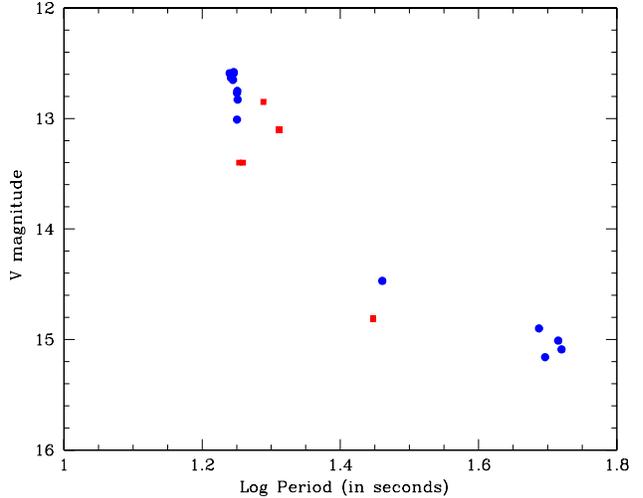,width=8.4cm}}}
  \caption{DNO periods as a function of the V magnitude of OY Car. The filled circles represent
this work, the filled squares are DNOs reported by Schoembs (1986) and Marsh \& Horne (1988).}
 \label{dno8fig5}
\end{figure}

    We can combine the DNOs listed in Schoembs, Marsh \& Horne and this work into two informative 
diagrams. A commonly plotted variation is the period-luminosity relationship, as seen in Fig.~\ref{dno8fig5}. 
The comparable plot for VW Hyi is given in fig. 8 of Paper I, for which we made the comment that 
from maximum down almost to quiescence the slope is typical of dwarf novae during decline; but then, 
in VW Hyi, there is a section where the period changes very rapidly without much drop in 
brightness. We found in VW Hyi (Paper IV) that this phase of rapid deceleration of the 
oscillations is also where frequency doubling and tripling occurs. In OY Car the same large 
reduction of slope in the $V - \log P_{\rm DNO}$ diagram is discernable, even though poorly observed, but 
is recognizable through direct comparison with VW Hyi.

\begin{figure}
\centerline{\hbox{\psfig{figure=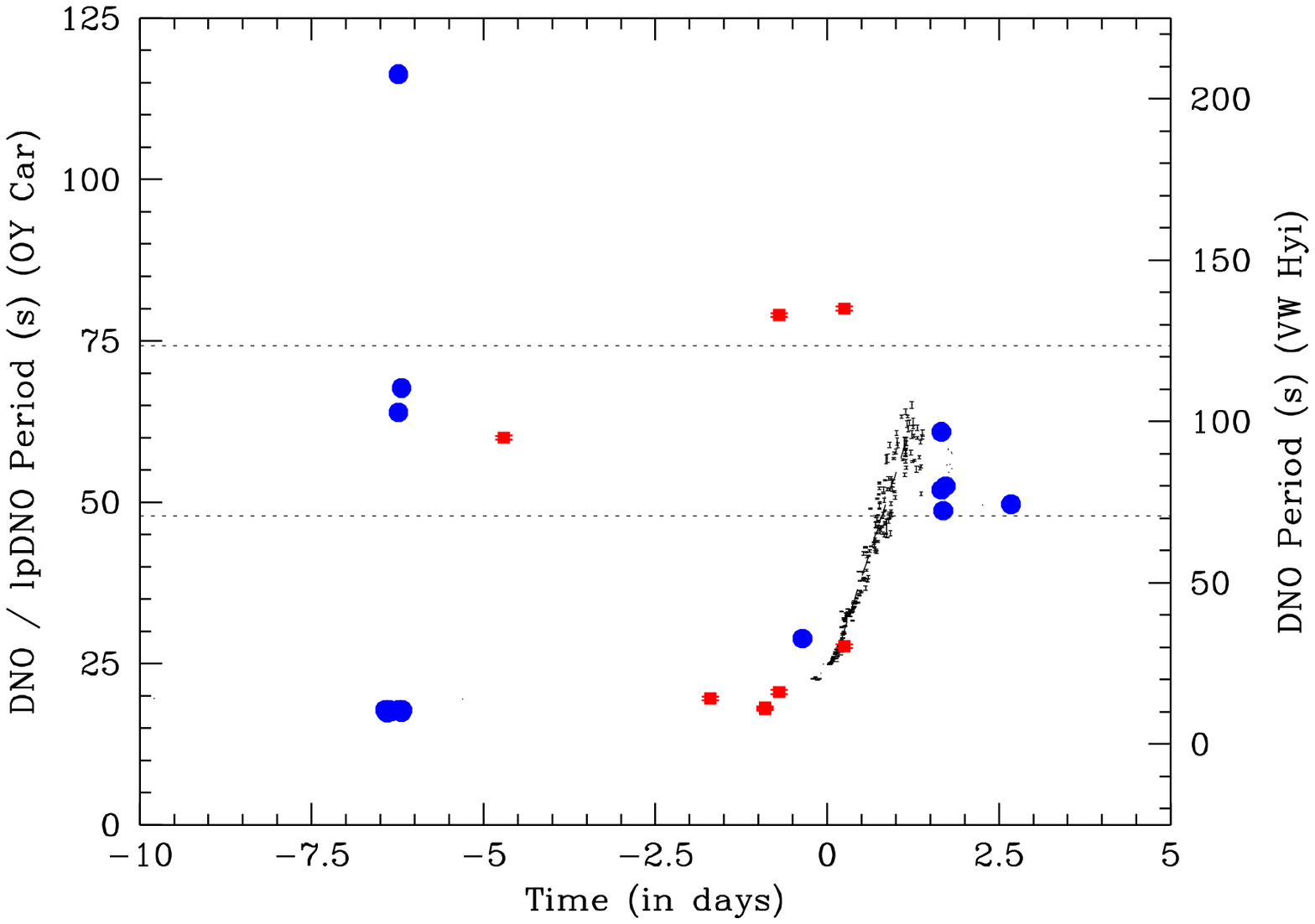,width=8.4cm}}}
  \caption{The evolution of DNO periods at the end of superoutbursts. The two horizontal dashed lines indicate
the range of the majority of (lp)DNOs observed during quiescence. The filled circles are from
this work, the filled squares are DNOs reported by Schoembs (1986) and Marsh \& Horne (1988). For comparison,
the implied fundamental DNO period of VW Hyi during the rapid deceleration phase is displayed by the small
dots; its scale is shown on the right-hand ordinate.}
 \label{dno8fig6}
\end{figure}

   The second diagram shows the evolution of DNO and lpDNO periods with time: Fig.~\ref{dno8fig6}. Again 
there is similarity to the behaviour of VW Hyi, but without evidence yet for frequency 
multiplication. The fundamental periods of the 
VW Hyi DNOs increase from 20.0 s at $T \sim 0$ to $\sim 90$ s at $T \sim 1.2$ d and then appear to stabilize at 
the latter period, which is similar to its lpDNOs (see Paper IV and fig. 2 of Warner \& Woudt 2009). 
In the case of OY Car the DNOs increase from 17.6 s and stabilize at $\sim 60$ s.

\begin{table*}
 \centering
  \caption{DNOs, lpDNOs and QPOs in OY Carinae.}
  \begin{tabular}{@{}lccrcccccc@{}}
\hline
 Run No. & O/Q$^{\dag}$ & HJD start  & Length         & \multicolumn{2}{c}{DNOs}  & \multicolumn{2}{c}{lpDNOs}    & \multicolumn{2}{c}{QPOs} \\
         &     & (2440000 +) & (s)            & \multicolumn{2}{c}{(periods in seconds)} & \multicolumn{2}{c}{(periods in seconds)} & \multicolumn{2}{c}{(period in seconds)}\\
         &     &[ Section ]  &                & \multicolumn{2}{c}{[amplitude in mmag]} & \multicolumn{2}{c}{[amplitude in mmag]} & \multicolumn{2}{c}{[amplitude in mmag]} \\
\hline
S6722    & O & 12672.36511  &  761           & \multicolumn{2}{c}{17.84 (0.05)}    & \multicolumn{2}{c}{--}       &        &   --    \\
         &   &   [ I ]      &                & \multicolumn{2}{c}{[5.4]}           &        &                     &        &         \\
         &   & 12672.37662  &  1326          & \multicolumn{2}{c}{17.56 (0.03)}    & \multicolumn{2}{c}{--}       &        &   --    \\
         &   &   [ II ]     &                & \multicolumn{2}{c}{[4.8]}           &        &                     &        &         \\
         &   & 12672.39202  &  1377          & \multicolumn{2}{c}{17.37 (0.02)}    & \multicolumn{2}{c}{--}       &        &   --    \\
         &   &   [ III ]    &                & \multicolumn{2}{c}{[7.4]}           &        &                     &        &         \\
         &   & 12672.42741  &  1260          & \multicolumn{2}{c}{17.82 (0.05)}    & \multicolumn{2}{c}{--}       & \vline &         \\
         &   &   [ IV ]     &                & \multicolumn{2}{c}{[3.1]}           &        &                     & \vline &         \\
         &   & 12672.44205  &  2325          & \multicolumn{2}{c}{17.61 (0.01)}    & \multicolumn{2}{c}{--}       & \vline & 278 (1) \\
         &   &   [ V ]      &                & \multicolumn{2}{c}{[5.4]}           &        &                     & \vline & [10.8]  \\
         &   & 12672.46907  &  1005          & \multicolumn{2}{c}{--}              & \multicolumn{2}{c}{--}       & \vline &         \\
         &   &   [ VI ]     &                &                            &        &        &                     & \vline &         \\
         &   &              &                &                            &        &        &                     &        &         \\
S6724    & O & 12672.54502  &  3445          & \multicolumn{2}{c}{17.80 (0.01)}    & 116.3 (0.9)  & 63.9 (1.2)    & \vline &         \\
         &   &    [ I ]     &                & \multicolumn{2}{c}{[4.7]}           & [9.3]        & [6.3]         & \vline &         \\
         &   & 12672.58501  &  1550          & \multicolumn{2}{c}{17.61 (0.02)}    & \multicolumn{2}{c}{--}       & \vline & 315 (1) \\
         &   &   [ II ]     &                & \multicolumn{2}{c}{[4.5]}           &             &                & \vline & [8.8]   \\
         &   & 12672.60301  &  945           & \multicolumn{2}{c}{17.44 (0.07)}    & \multicolumn{2}{c}{67.7 (0.5)} & \vline &       \\
         &   &   [ III ]    &                & \multicolumn{2}{c}{[3.3]}           & \multicolumn{2}{c}{[9.8]}      & \vline &       \\
         &   & 12672.61401  &  940           & \multicolumn{2}{c}{17.80 (0.04)}    & \multicolumn{2}{c}{--}       &        &   --    \\
         &   &   [ IV ]     &                & \multicolumn{2}{c}{[6.1]}           &        &                     &        &         \\
         &   &              &                &                            &        &        &                     &        &         \\
S6748    & O & 12678.42970  &  1790          & \multicolumn{2}{c}{28.89 (0.06)}    & \multicolumn{2}{c}{--}       &        &   --    \\
         &   &              &                & \multicolumn{2}{c}{[11.6]}          &        &                     &        &         \\
         &   &              &                &                            &        &        &                     &        &         \\
S6754    & Q & 12680.42688  &  990           & 42.85 (0.22) & 48.91 (0.31)         & \multicolumn{2}{c}{--}       & \vline & {\it 346} \\
         &   &   [ I ]      &                & [16.8]       & [15.8]               &        &                     & \vline &         \\
         &   & 12680.44714  &  2395          & 51.90 (0.20) & 60.89 (0.27)         & \multicolumn{2}{c}{--}       & \vline & {\it 352} \\
         &   &   [ II ]     &                & [12.7]       & [12.8]               &        &                     & \vline &         \\
         &   & 12680.47491  &  2441          & \multicolumn{2}{c}{48.68 (0.14)}    & \multicolumn{2}{c}{--}       & \vline & 305 (1) \\
         &   &   [ III ]    &                & \multicolumn{2}{c}{[12.1]}          &        &                     & \vline & [22.0]  \\
         &   & 12680.51156  &  2266          & \multicolumn{2}{c}{52.48 (0.23)}    & \multicolumn{2}{c}{--}       & \vline & 571 (2) \\
         &   &   [ IV ]     &                & \multicolumn{2}{c}{[11.1]}          &        &                     & \vline & [34.2]  \\
         &   &              &                &              &                      &        &                     &        &         \\
S6755    & Q & 12681.46269  &  2106          & \multicolumn{2}{c}{49.68 (0.19)}    & \multicolumn{2}{c}{--}       &  &  --     \\
         &   &              &                & \multicolumn{2}{c}{[11.9]}          &        &                     &  &         \\
         &   &              &                &              &                      &   &                &        &         \\
S2966    & Q &  5053.32008  &  2320          & \multicolumn{2}{c}{--}              & \multicolumn{2}{c}{104.4 (0.4)}  &  &   --    \\
         &   &              &                &              &                      & \multicolumn{2}{c}{[24.6]}       &  &         \\
         &   &              &                &              &                      &   &                &        &         \\
S2982    & Q &  5108.21176  &  3760          & \multicolumn{2}{c}{--}              & \multicolumn{2}{c}{130.4 (0.7)}  &  &   --    \\
         &   &              &                &              &                      & \multicolumn{2}{c}{[16.6]}       &  &         \\
         &   &              &                &              &                      &   &                &        &         \\
S6485    & Q & 12321.32548  &  4360          & \multicolumn{2}{c}{--}              & \multicolumn{2}{c}{66.99 (0.20)} &  &  --     \\
         &   &    [ I ]     &                &              &                      & \multicolumn{2}{c}{[16.9]}       &  &         \\
         &   & 12321.38501  &  1120          & \multicolumn{2}{c}{--}              & \multicolumn{2}{c}{74.20 (0.61)} &  &  --     \\
         &   &    [ II ]    &                &              &                      & \multicolumn{2}{c}{[25.5]}       &  &         \\
         &   & 12321.44501  &   860          & \multicolumn{2}{c}{--}              & 61.39 (0.37)  & 74.26 (0.55)     &  & {\it 354} \\
         &   &    [ III ]   &                &              &                      & [17.5]        & [16.6]           &  &         \\
         &   & 12321.48002  &  1898          & \multicolumn{2}{c}{--}              & 56.92 (0.20)  & 69.71 (0.23)     &  & {\it 310} \\
         &   &    [ IV ]    &                &              &                      & [9.0]         & [11.6]           &  &         \\
         &   & 12321.54198  &  1120          & \multicolumn{2}{c}{--}              & \multicolumn{2}{c}{48.35 (0.23)} & \vline  & 585 (3) \\
         &   &    [ V ]     &                &              &                      & \multicolumn{2}{c}{[16.9]}       & \vline  & [32.6]  \\
         &   & 12321.55500  &  1728          & \multicolumn{2}{c}{--}              & \multicolumn{2}{c}{50.71 (0.15)} & \vline  &         \\
         &   &    [ VI ]    &                &              &                      & \multicolumn{2}{c}{[14.2]}       & \vline  &         \\
         &   &       &                &              &                &   &                &        &         \\
S6488    & Q & 12323.50036  &  1088          & \multicolumn{2}{c}{--}              & \multicolumn{2}{c}{60.52 (0.59)} &  &  --     \\
         &   &    [ I ]     &                &              &                      & \multicolumn{2}{c}{[26.3]}       &  &         \\
         &   & 12323.51303  &  1120          & \multicolumn{2}{c}{--}              & \multicolumn{2}{c}{47.84 (0.16)} &  &  --     \\
         &   &    [ II ]    &                &              &                      & \multicolumn{2}{c}{[29.3]}       &  &         \\
\hline
\end{tabular}
\newline
{\footnotesize Notes: Uncertainties in the periods are quoted after the periods in parentheses; 
$^{\dag}$ O/Q = Outburst/Quiescence.}
\label{dno8tab2}
\end{table*}

\subsection{Observations during deep quiescence}

Schoembs (1986) was the first to observe DNOs in OY Car at quiescence -- he found oscillations 
at 55 s and at 122 s in a light curve obtained 8 days after the end of a superoutburst. For our 
observations, returning to Tables~\ref{dno8tab1} and \ref{dno8tab2}, of 10 runs of sufficient length 
and quality to detect oscillations, made in quiescence well away from outbursts, we have three positive 
identifications. Run S2966 shows a very strong oscillation at 104 s for about 40 min; but it 
is in runs S6485 and S6488, made on consecutive nights, that we have a more informative haul. 
We have already illustrated a short section of S6488, showing large amplitude oscillations 
at $\sim 48$ s directly visible in the light curve (Paper III); here we analyse these quiescent 
DNOs in detail. An aspect that is conducive to detecting rapid oscillations is that the 
quiescent light curves of OY Car have unusually low flickering activity.

\begin{figure}
\centerline{\hbox{\psfig{figure=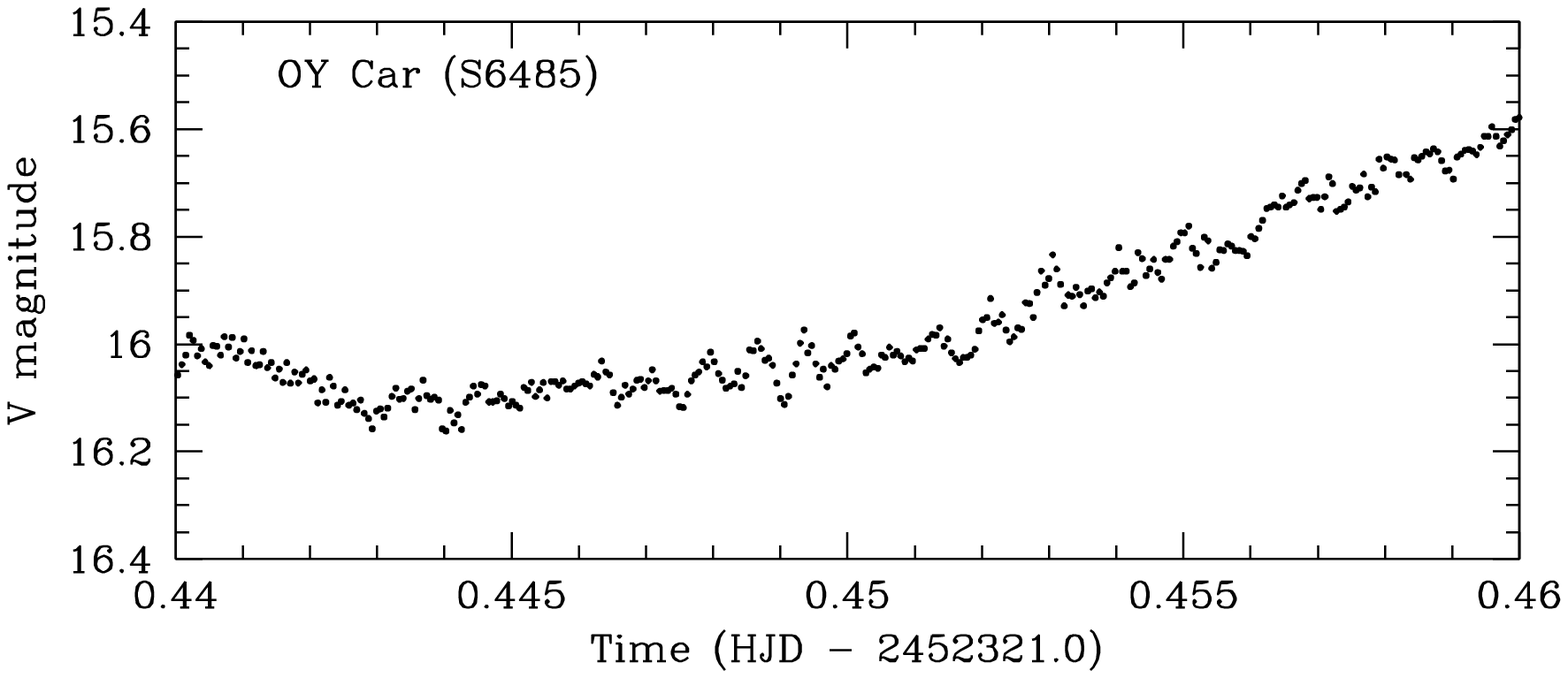,width=8.4cm}}}
  \caption{A section of the light curve of run S6485.}
 \label{dno8fig7}
\end{figure}

\begin{figure}
\centerline{\hbox{\psfig{figure=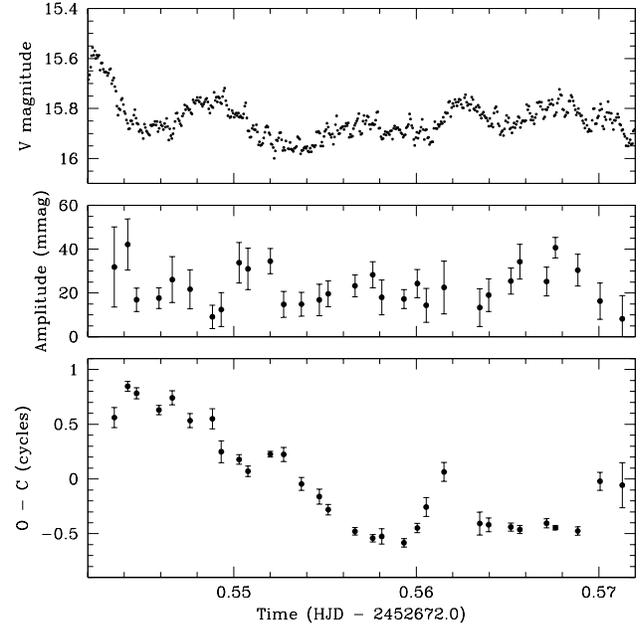,width=8.4cm}}}
  \caption{The $A - \phi$ diagram for run S6485, relative to a fixed period of 50.71 s; phase 
variations are shown in the lower panel, amplitude variations are displayed in the middle panel 
and the light curve is shown in the top panel.}
 \label{dno8fig8}
\end{figure}

  For these runs S6485 and S6488 Table~\ref{dno8tab2} shows occasional simultaneous presence of two DNOs, 
separated by $\sim 12$ s with beat period $\sim 330$ s, and also a tendency to switch between two similarly 
spaced periods. This is very similar to the simultaneous or alternating presence of direct 
and reprocessed beams mentioned in Section 1 and discussed in Paper VI. Therefore in OY Car 
we see, for the first time, an aspect of outburst behaviour also present in quiescence.  
Fig.~\ref{dno8fig7} shows a section of S6485 in which DNOs at 61 s and 74 s are present with sufficient 
amplitude that the beats between them can be seen. Fig.~\ref{dno8fig8} shows an $A - \phi$ diagram for 
part of run S6485 in which the abrupt shift from one DNO period to another is seen as a change of 
slope in the phases.

\begin{figure}
\centerline{\hbox{\psfig{figure=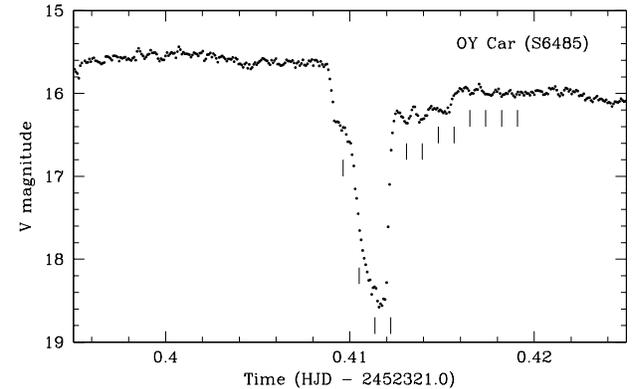,width=8.4cm}}}
  \caption{A section of the light curve of run S6485. Minima of the 74-s lpDNO are marked by vertical 
bars.}
 \label{dno8fig9}
\end{figure}

   Fig.~\ref{dno8fig9} shows the 74 s lpDNO emerging from eclipse. It appears after the central 
part of the disc has been uncovered but before the bright spot has reappeared, which is consistent with 
the lpDNO source being close to the primary. Fig.~\ref{dno8fig9} is very similar to the run 2290 eclipse of 
HT Cas shown in fig 8 of Patterson (1981), which was used to deduce that the flickering source 
in HT Cas is located near the centre of the disc -- we suggest that it was not flickering but 
the $\sim 100$ s lpDNO in HT Cas that was active at that time.

  The last two sections of S6485 show a QPO with period 585 s, which is roughly twice the period 
inferred from the double DNOs in the earlier part of the run. A similar effect was seen in run 
S6754 (Table~\ref{dno8tab2}). Switches between a QPO and its harmonic are commonly seen in VW Hyi 
(figure 10 of paper IV).

   Despite the various similarities between OY Car and VW Hyi we found no DNOs in the latter 
during quiescence to much lower amplitude limits than those we have observed in OY Car 
(table 1 of Paper I).

\subsection{DNOs during eclipse}

\begin{figure}
\centerline{\hbox{\psfig{figure=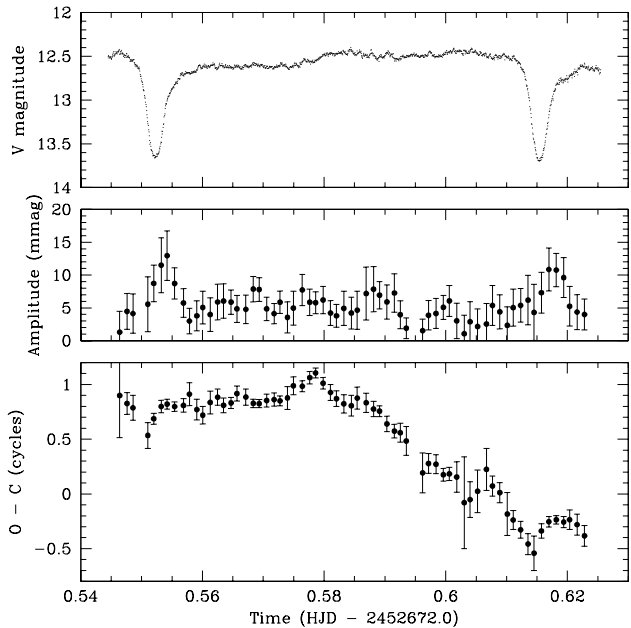,width=8.4cm}}}
  \caption{The $A - \phi$ diagram for run S6724, relative to a fixed period of 17.77 s; phase 
variations are shown in the lower panel, amplitude variations are displayed in the middle panel 
and the light curve is shown in the top panel.}
 \label{dno8fig10}
\end{figure}

\begin{figure}
\centerline{\hbox{\psfig{figure=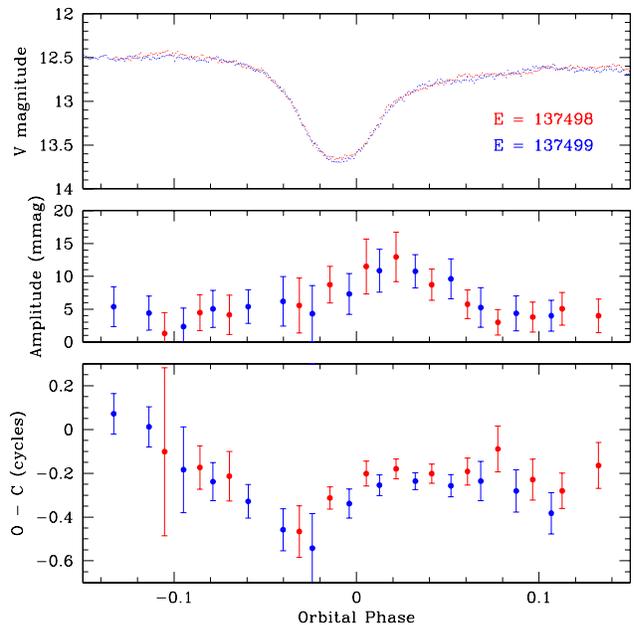,width=8.4cm}}}
  \caption{The $A - \phi$ diagram for run S6724, relative to a fixed period of 17.77 s folded
on the orbital ephemeris of Greenhill et al.~(2006); cycle numbers on Greenhill's ephemeris
are given in the top panel.}
 \label{dno8fig11}
\end{figure}

      Schoembs (1986) found that the amplitude of DNOs in two of his light curves obtained near 
supermaximum peaked somewhat after eclipse minimum. The phase variations did not show any 
definite variation through eclipse, but the DNO periods were not sufficiently stable to give 
reliable results. Our superouburst run S6724 is of high quality and possesses two eclipses that 
are suitable for analysis. The light curve was smoothed by subtracting a 15 bin running average 
before determining the $A - \phi$ diagram. The basic light curve and the amplitudes and phases of the 
DNOs are shown in Fig.~\ref{dno8fig10}. As can be seen from the phase diagram, there were changes of DNO 
period through the run (see Table~\ref{dno8tab2}); these prevent confident detection of systematic phase 
shifts through eclipse, but, as previously found by Schoembs, there are maxima in the DNO 
oscillation amplitudes well after eclipse minimum.  In Fig.~\ref{dno8fig11} we show the $A - \phi$ diagrams 
in the vicinity of the two eclipses, folded on orbital phase. Schoembs noted that in his runs maximum 
DNO amplitude corresponded to maximum of superhumps, but that may have been only a coincidence 
because this does not appear to be the case in our run. It is possible that the offset of pulse 
amplitude relative to mid-eclipse is caused by a strong reprocessing signal coming from the 
thickened disc in the vicinity of the impact of the accretion stream, as has recently been 
found for DQ Her (Saito \& Baptista 2009).

   Our analysis of the quiescent DNOs during eclipse, which is made difficult because of variable 
periods of the lpDNOs, shows no systematic phase changes and in particular no peak amplitude 
after mid-eclipse as is seen in outburst (Fig.~\ref{dno8fig11}).

\section{Other oscillations in quiescence}

\subsection{Other DNOs observed in quiescence}

In addition to the quiescent DNOs or lpDNOs discussed in the previous Section there are 
reports of hard X-Ray DNOs in other quiescent dwarf novae. Cordova \& Mason (1984) and Eracleous, 
Patterson \& Halpern (1991), from the same half an hour length of data, find a period at 21.83 s 
in HT Cas. Although intrinsically of weak statistical significance this is close to the 
20.2 -- 20.4 s range of DNOs observed during outburst (Patterson 1981) and therefore gains 
credibility. Pandel, Cordova \& Howell (2003) found a 62 s modulation in quiescence in VW Hyi, with an 
amplitude of 6\% and of similar strengths in hard and soft X-Rays. From the width of the signal 
in the Fourier transform they deduced a coherence length of only 2 cycles, but the spectral 
broadening is certainly due to a steady change of period and alternating direct and reprocessed 
contributions of a far more coherent DNO (Paper VI). U Gem showed periods of 121 and 135 s 
(table 3(a) of Cordova \& Mason (1984)) in quiescence, the beat period between which is 
1166 s, which is almost exactly twice the 585 s QPO period seen in hard X-rays during an outburst 
of U Gem (Cordova \& Mason 1984), suggesting that the shorter periods are lpDNOs and that 
frequency doubling (Section 1) of the QPO had taken place. This is also consistent with $\sim 25$ s 
DNOs observed in soft X-Rays during outbursts of U Gem (Mason et al.~1988). One other claim is 
a statistically significant 33.9 s DNO in SU UMa (Eracleous et al.~1991), which has had no 
optically detected DNOs even during outburst to compare with it.

\subsection{QPOs in quiescence}

Although rare, there are previous observations of optical QPOs in the quiescent state of 
some dwarf novae. The earliest was of 1200 -- 1800 s modulation in SS Aur with amplitude up 
to 0.5 mag (Tovmassian 1987, 1988). Recently RAT J1953+1859 was initially discovered to be 
a CV through the presence of 0.3 mag QPOs at $\sim 1200$ s, and later found to outburst as a dwarf 
nova (Ramsay et al.~2009). Between these dates a QPO at 343 s and amplitude 0.03 mag, with a 
possible orbital sideband, was found in V893 Sco (Bruch, Steiner \& Gneiding 2000; Paper III) 
and a QPO at 190 s in WX Hyi, reported in Paper V, is similar to one seen in outburst. These 
latter are probably lpDNOs. QPOs at $\sim 600 - 1000$ s were found to be quite common in VW Hyi at 
quiescence (Paper I).

    There are also two observations of quiescent QPOs made in the X-ray region: 290 s in 
AB Dra (Cordova \& Mason 1984), which is possibly also an lpDNO, and a pair at 2193 
and 3510 s in OY Car (Ramsay et al.~2001; Hakala \& Ramsay 2004).

\section{Discussion}

  Our observations of OY Car during outburst show that it shares many behavioural properties 
with the more comprehensively studied VW Hyi. As with VW Hyi its DNOs can be followed into 
quiescence for a few days after outburst, but in addition it can show occasional DNOs in 
quiescence at any time between outbursts. Our analysis indicates that the quiescent DNOs in 
many ways resemble outburst DNOs, and therefore require similar physical explanations.

    The notion that a LIMA model is applicable in the low $\dot{M}$ state is supported by the 
conclusion of Wheatley \& West (2003) from hard X-Ray observations of OY Car during quiescence. 
They find that eclipse of the emitting source is of much shorter duration than that of the white 
dwarf and is displaced such that the X-Rays evidently arise in a small region at high latitude 
on the surface of the primary. This is what is expected from magnetically controlled accretion 
onto an underlying inclined dipole field of the primary, rather than onto a stronger field 
generated during outburst by the differentially rotating equatorial belt.  

   We have presented observations that suggest that, although rarer than some of those seen 
during outburst, DNOs during quiescence are present in a number of dwarf novae, should be 
searched for more assiduously, and need to be accommodated in any general model of CV oscillations.

\section*{Acknowledgments}

Both authors acknowledge support from the University of Cape Town and from the National 
Research Foundation of South Africa. PAW in addition thanks the School of Physics and 
Astronomy at Southampton University for financial assistance. We kindly acknowledge the use
of the AAVSO data archive of OY Car which was used to generate the average superoutburst
profile. We furthermore thank observers at UCT who contributed to the data archive of OY Car 
over the years.


\begin{thebibliography}{99}
\bibitem{bru02}   Bruch A., Steiner J.E., Gneiding C.D., 2000, PASP, 112, 237
\bibitem{che97}   Cheng F.H., Sion E.M., Horne K., Hubeny I., Huang M., Vrtilek S.D., 1997, AJ, 114, 1165
\bibitem{cor84}   Cordova F.A., Mason K.O., 1984, MNRAS, 206, 879
\bibitem{erac91}  Eracleous M., Patterson J., Halpern J.P., 1991, ApJ, 370, 330
\bibitem{godo04}  Godon P., Sion E.M., Cheng F.H., Szkody P., Long K.S., Froning C.S., 2004, ApJ, 612, 429
\bibitem{gre06}   Greenhill J.G., Hill K.M., Dieters S., Fienberg K., Howlett M., Meijers A.,
  Munro A., Senkbeil C., 2006, MNRAS, 372, 1129
\bibitem{haka04}  Hakala P., Ramsay G., 2004, A\&{A}, 416, 1047
\bibitem{mars88}  Marsh T.R., Horne K., 1988, MNRAS, 299, 921
\bibitem{maso88}  Mason K.O., Cordova F.A., Watson M.G., King A.R., 1988, MNRAS, 232, 779
\bibitem{odo95}   O'Donoghue D., 1995, Balt. Astr., 4, 517
\bibitem{p78}     Paczynski B., 1978, in Zytkow A., ed., Nonstationary Evolution of Close Binaries.
  Polish Scientific Publ., Warsaw, p.~89
\bibitem{pan03}   Pandel D., Cordova F., Howell S.B., 2003, MNRAS, 346, 1241
\bibitem{pan05}   Pandel D., Cordova F., Mason K.O., Priedhorsky W.C., 2005, ApJ, 626, 396
\bibitem{pat81}   Patterson J., 1981, ApJS, 45, 517
\bibitem{piro09}  Piro A.L., Bildsten L., 2004, ApJ, 616, L155
\bibitem{pre06}   Pretorius M.L., Warner B., Woudt P.A., 2006, MNRAS, 368, 361 (paper V)
\bibitem{rams01}  Ramsay G., et al., 2001,  A\&{A}, 365, 288
\bibitem{rams09}  Ramsay G., et al., 2009, MNRAS, in press (arXiv:0904.4211)
\bibitem{saito09} Saito R.K., Baptista R., 2009, ApJ, 693, L16.
\bibitem{scho86}  Schoembs R., 1986, A\&{A}, 158, 233
\bibitem{sion99}  Sion E.M., 1999, PASP, 111, 532
\bibitem{sion02}  Sion E.M., Urban J., 2002, ApJ, 572, 456
\bibitem{sion04}  Sion E.M., Cheng F.H., Godon P., Urban J.O., Szkody P., 2004, AJ, 128, 1834
\bibitem{tovm87}  Tovmassian G.G., 1987, Astrofiz, 27, 231
\bibitem{tovm88}  Tovmassian G.G., 1988, Adv. Sp. Res., 8, 329
\bibitem{urba06}  Urban J.O., Sion E.M., 2006, ApJ, 642, 1029
\bibitem{vogt79}  Vogt N., 1979, Messenger, 17, 39
\bibitem{war95}   Warner B., 1995, Cataclysmic Variable Stars, Cambridge Univ. Press, Cambridge
\bibitem{war04}   Warner B., 2004, PASP, 116, 115
\bibitem{wapr08}  Warner B., Pretorius M.L., 2008, MNRAS, 383, 1469 (paper VI)
\bibitem{wawo02}  Warner B., Woudt P.A., 2002, MNRAS, 335, 84 (paper II)
\bibitem{wawo05}  Warner B., Woudt P.A., 2005, in Hameury J.-M., Lasota J.-P., eds., ASP Conf. Ser. Vol. 330,
  The Astrophysics of Cataclysmic Variables and Related Objects. Astron. Soc. Pac., San Francisco, p. 227
\bibitem{wawo06}  Warner B., Woudt P.A., 2006, MNRAS, 367, 1562 (paper IV)
\bibitem{wawo08}  Warner B., Woudt P.A., 2008, in Axelsson M., ed., AIP Conf. Proc. Vol. 1054, 
  COOL DISCS, HOT FLOWS: The Varying Faces of Accreting Compact Objects. American Institute of 
  Physics, New York, p. 101
\bibitem{wawo09}  Warner B., Woudt P.A., 2009, in Soonthornthum B., et al., eds., ASP Conf. Ser. Vol. 404, 
  The Eighth Pacific Rim Conference on Stellar Astrophysics: A Tribute to Kam Ching Leung. 
  Astron. Soc. Pac., San Francisco, p. 45
\bibitem{wwp03}   Warner B., Woudt P.A., Pretorius M.L., 2003, MNRAS, 334, 1193 (paper III)
\bibitem{wewe03}  Wheatley P.J., West R.G., 2003, MNRAS, 345, 1009
\bibitem{wowa02}  Woudt P.A., Warner B., 2002, MNRAS, 333, 411 (paper I)
\bibitem{woea09}  Woudt P.A., Warner B., O'Donoghue D., Buckley D.A.H., Still M., Romero-Colmenero E.,
  V\"ais\"anen P., 2009, MNRAS, accepted (paper VIII)
\end{thebibliography}
\end{document}